# Thermodynamics of Ferroelectric and Optical Properties in $KNbO_3$

Aiden Ross*, Venkatraman Gopalan, and Long-Qing Chen

Potassium niobate ($KNbO_3$) is a prototypical perovskite ferroelectric with large electro-optic and nonlinear optical responses, high optical damage thresholds and a rich sequence of temperature-driven phase transformations, making it a promising platform for tunable photonic devices. In this work, we develop a thermodynamic model for the coupled ferroelectric and optical properties of $KNbO_3$. By separating the total polarization into lattice and electronic contributions, the model provides a unified description of both the anisotropic ferroelectric and optical properties. The thermodynamic coefficients are determined by fitting to experimental measurements of the spontaneous polarization, dielectric susceptibilities, lattice parameters, and refractive indices. Without any further fitting, the model quantitatively predicts the temperature dependence of the electro-optic and piezoelectric coefficients in close agreement with experimental measurements. By utilizing the electronic polarization equation of motion, the model further captures the optical dispersion in the near infrared to visible spectrum. This work provides the thermodynamic foundation for future studies of coupled ferroelectric and optical phenomena in $KNbO_3$.

[1]Department of Materials Science and Engineering and Materials Research Institute, The Pennsylvania State University, University Park, PA 16802

* Email Address: amr8057@psu.edu

## I. INTRODUCTION

Potassium niobate (KNbO$_3$) is a prototypical perovskite ferroelectric with a large electro-optic response [1], high nonlinear optical coefficients [2], and a high threshold for optical damage, making it a promising material for optical modulators, frequency conversion devices, and emerging integrated photonic platforms. Importantly, KNbO$_3$ is highly tunable through composition [3,4], external electric fields, epitaxial strain [5,6] and domain engineering [3,5], offering a versatile platform to control light-matter interactions.

The optical properties of KNbO$_3$ are closely linked to its ferroelectric behavior [7] . Upon cooling, KNbO$_3$ undergoes a series of structural phase transformations from cubic to tetragonal, orthorhombic and rhombohedral, and each of the phases is characterized by a different spontaneous polarization direction and corresponding changes in the symmetry of the optical properties (e.g. refractive indices, electro-optic tensor, second harmonic generation tensor). In addition, when grown as a strained thin film, additional low-symmetry monoclinic phases that are absent in the bulk phase diagram appear [5,6]. These ferroelectric phases and phase-transitions provide opportunities to engineer new optical functionalities, but significant challenges remain to theoretically understand and predict their optical responses.

Conventional thermodynamic approaches based on Landau-Ginzburg-Devonshire theory have been employed to study the ferroelectric phase transitions in KNbO$_3$ [8]. These models are parameterized utilizing experimental data including the spontaneous polarization, the spontaneous strain, and the dielectric susceptibility and allow one to consistently calculate the properties of KNbO$_3$ across a broad temperature range and under different elastic boundary conditions. These thermodynamic models nevertheless fail to accurately capture the optical properties in the near-infrared to visible spectrum since they do not separate the lattice and electronic contributions to the total polarization [9].

Building upon our recent advance in integrating the ferroelectric and optical properties into a unified thermodynamic framework [9], here we develop a thermodynamic model for the ferroelectric and optical properties of KNbO$_3$. We fit the thermodynamic model using experimental data on the spontaneous polarization [10–13], refractive indices [14,15], and dielectric susceptibilities [16,17], and then use it to predict the temperature-dependent electro-optic effect [1,18] and piezoelectric effect [19–21]. This work provides the thermodynamic foundation for predicting the temperature-, strain- and electric field- dependent properties of KNbO$_3$ and can be extended to inhomogeneous ferroelectric systems containing domain structures using the phase-field method.

## II. THEORY

Following from the thermodynamic theory of optical properties in ferroelectrics [9], we may describe KNbO$_3$ by separating the lattice and electronic contributions to the total polarization. Here the lattice polarization ($P^L$) is the contribution that arises solely from the displacement of ions in the crystal lattice, and the electronic polarization ($P^e$) corresponds to field induced displacements of the centers of charge in the electronic orbitals (the Wannier centers) from their zero-field equilibrium positions. Using this separation, we describe the total free-energy density as the summation of the lattice ($f^L$) , electronic ($f^e$), lattice-electronic coupling ($f^{L-e}$) , elastic ($f^{elas}$), and electrostatic energies ($f^{elec}$),

$$f(T, P_i^L, P^e, \sigma_{ij}, E_i) = f^L(T, P_i^L) + f^e(T, P_i^e) + f^{L-e}(P_i^L, P_i^e) + f^{elas}(P_i^L, P_i^e, \sigma_{ij}) + f^{elec}(P_i^L, P_i^e, E_i) \qquad (1)$$

where $\sigma_{ij}$ is the stress tensor and $E_i$ is the electrical field.

Here $f^L(P_i^L)$ describes the intrinsic stability of the lattice polarization compared to the high symmetry phase ($m\bar{3}m$) as a Taylor expansion of the polarization about the high symmetry phase, this is equivalent to the Landau free energy density

$$f^L(T, P_i^L) = a_{ij}(T)P_i^L P_j^L + a_{ijkl}P_i^L P_j^L P_k^L P_l^L + a_{ijklmn}P_i^L P_j^L P_k^L P_l^L P_m^L P_n^L + \qquad (2)$$
$$a_{ijklmnop}P_i^L P_j^L P_k^L P_l^L P_m^L P_n^L P_o^L P_p^L,$$

where $a_{ij}$, $a_{ijkl}$, $a_{ijklmn}$, and $a_{ijklmnop}$ are the dielectric stiffness coefficients measured under stress-free conditions. We employ an 8th order free energy density function by adjusting the coefficients from Liang et. al. [8] to better fit the temperature-dependent polarization, and dielectric susceptibilities and describe the effects of quantum fluctuations occurring at cryogenic temperatures [22].

The intrinsic free energy density of the electronic polarization, in the absence of the lattice polarization, is described by

$$f^e(T, P_i^e) = \frac{1}{2\epsilon_0} B_{ij}^{e,ref}(T) P_i^e P_j^e + \frac{1}{24\epsilon_0} g_{ijkl}^{ee} P_i^e P_j^e P_k^e P_l^e, \qquad (3)$$

where $B_{ij}^{ref}(T)$ is related to the refractive index of the cubic phase and $g_{ijkl}^{ee}$ is related to the optical third harmonic generation. The coupling energy density between the lattice and electronic polarization which determines the electro-optic effect is given by

$$f^{L-e}(P_i^L, P_i^e) = \frac{1}{2\epsilon_0} g_{ijkl}^{LL} P_i^e P_j^e P_k^L P_l^L + \frac{1}{6\epsilon_0} g_{ijkl}^{Le} P_i^e P_j^e P_k^e P_l^L, \qquad (4)$$

where $g_{ijkl}^{LL}$ couples the lattice polarization to the curvature of the electronic polarization energy landscape responsible for the anisotropic refractive indices and $g_{ijkl}^{Le}$ couples the lattice polarization to the anharmonicity of the electronic polarization energy landscape giving rise to first order nonlinearities including second harmonic generation and sum/difference frequency generation. For simplicity in this manuscript, we will only parameterize $B_{ij}^{ref}$ and $g_{ijkl}^{LL}$ and ignore the contributions of $g_{ijkl}^{Le}$ and $g_{ijkl}^{ee}$.

The elastic energy is given by

$$f^{elas}(P_i^L, P_i^e, \sigma_{ij}) = -\frac{1}{2} s_{ijkl}\sigma_{ij}\sigma_{kl} - Q_{ijkl}\sigma_{ij} P_k P_l - \frac{1}{2\epsilon_0} \pi_{ijkl}\sigma_{ij} P_k^e P_l^e, \qquad (5)$$

where $s_{ijkl}$ is the elastic compliance tensor, $Q_{ijkl}$ is the electrostrictive tensor, $\pi_{ijkl}$ is the piezo-optic tensor of the parent phase and where $\epsilon_0$ is the vacuum permittivity.

The electrostatic energy is given by

$$f^{elec} = -E_i P_i^L - E_i P_i^e - \frac{1}{2}\epsilon_0 \delta_{ij} E_i E_j, \qquad (6)$$

where $\delta_{ij}$ is the Kronecker delta. Here we use the vacuum permittivity and assume that the electronic polarization fully accounts for the background dielectric constant [23].

Using this thermodynamic energy density function, one can evaluate the ferroelectric and optical properties in the thermodynamic limit (nondispersive and no absorption). The equilibrium lattice and electronic polarizations are found by solving the following coupled equations

$$\left(\frac{\partial f}{\partial P_i^L}\right)_{T,\sigma,P^e} = 0, \left(\frac{\partial f}{\partial P_i^e}\right)_{T,\sigma,P^L} = 0. \qquad (7)$$

The lattice dielectric susceptibility is then given by

$$\chi^L_{ij} = \frac{1}{\epsilon_0}\left(\frac{\partial P^L_i}{\partial E_j}\right)_{T,\sigma} = \frac{1}{\epsilon_0}\left(\frac{\partial^2 f}{\partial P^L_i \partial P^L_j}\right)^{-1}_{T,\sigma}, \tag{8}$$

and the electronic dielectric susceptibility is

$$\chi^e_{ij} = (n^2)_{ij} - \delta_{ij} = (B^e_{ij})^{-1} = \frac{1}{\epsilon_0}\left(\frac{\partial P^e_i}{\partial E_j}\right)_{T,\sigma} = \frac{1}{\epsilon_0}\left(\frac{\partial^2 f}{\partial P^e_i \partial P^e_j}\right)^{-1}_{T,\sigma}, \tag{9}$$

where $n$ is the refractive index and $B^e_{ij}$ is the electronic dielectric stiffness. We can also derive the index ellipsoid or optical dielectric stiffness by rearranging equation 9

$$B_{ij} = (n^2)^{-1}_{ij} = B^e_{ik}[(B^e + \mathbb{I})^{-1}]_{kj}, \tag{10}$$

where $\mathbb{I}$ is the identity tensor.

We may also calculate the electro-optic effect by considering the lattice and electronic contributions to the electro-optic effect

$$r_{ijk} = \left(\frac{\partial B_{ij}}{\partial E_k}\right)_{T,\sigma} = r^L_{ijk} + r^e_{ijk} \tag{11}$$

where $r^L_{ijk}$ is the lattice contribution to the electro-optic effect, and $r^e_{ijk}$ is the electronic contribution to the electro-optic effect. Using the chain rule, each contribution can be decomposed into a polar-optic effect and a dielectric susceptibility,

$$r^L_{ijk} = \left(\frac{\partial B_{ij}}{\partial P^L_m}\right)_{T,\sigma}\left(\frac{\partial P^L_m}{\partial E_k}\right)_{T,\sigma} = \epsilon_0 f^L_{ijm}\chi^L_{mk} \tag{12}$$

$$r^e_{ijk} = \left(\frac{\partial B_{ij}}{\partial P^e_m}\right)_{T,\sigma}\left(\frac{\partial P^e_m}{\partial E_k}\right)_{T,\sigma} = \epsilon_0 f^e_{ijm}\chi^e_{mk} \tag{13}$$

where $f^L_{ijk}$, and $f^e_{ijk}$ are the polar-optic effects for the lattice and electronic polarization respectively. A more detailed derivation which includes the thermo-optic effect and piezo-optic effect is given in [9].

To account for the optical dispersion and loss, we can utilize the polarization equation of motion

$$\mu^e_{ij}\frac{\partial^2 P^e_j}{\partial t^2} + \gamma^e_{ij}\frac{\partial P^e_j}{\partial t} = \frac{\delta F}{\delta P^e_i} \tag{14}$$

where $\mu^e_{ij}$ is the electronic polarization effective mass, and $\gamma^e_{ij}$ is the electronic polarization damping coefficient. We may solve equation 14 using a perturbation expansion [24] which yields

$$\widetilde{\chi^e_{ij}}(\omega) = \tilde{n}^2_{ij} - \delta_{ij} = \left[B^e_{ij} - \epsilon_0(i\omega\gamma^e_{ij} + \omega^2\mu^e_{ij})\right]^{-1}, \tag{15}$$

where $\tilde{n}$ is the refractive index. By rearranging equation 15 we also find the frequency dependent optical dielectric stiffness,

$$B_{ij}(\omega) = [B^e - \epsilon_0(i\omega\gamma^e + \omega^2\mu^e)]_{ik}[(B^e - \epsilon_0(i\omega\gamma^e + \omega^2\mu^e) + \mathbb{I})^{-1}]_{kj}. \tag{16}$$

Ignoring absorption, the frequency dependent electro-optic effect may then be calculated via

$$r^L_{ijk}(\omega) = \left(\frac{\partial B_{ij}(\omega)}{\partial P^L_m}\right)_{T,\sigma}\left(\frac{\partial P^L_m}{\partial E_k}\right)_{T,\sigma} = f^L_{ijm}(\omega)\chi^L_{mk} \tag{17}$$

where

$$f^L_{ijk} = \left(\frac{\partial B_{ij}(\omega)}{\partial P^L_k}\right)_{T,\sigma} = \left(\frac{\partial B_{ij}(\omega)}{\partial B^e_{mn}}\right)_{T,\sigma}\left(\frac{\partial B^e_{mn}}{\partial P^L_k}\right)_{T,\sigma}$$

$$= [(B^e - \epsilon_0\omega^2\mu^e + \mathbb{I})^{-1}]_{mi}[(B^e - \epsilon_0\omega^2\mu^e + \mathbb{I})^{-1}]_{nj}\left[2g^{LL}_{mnkl}P^L_l + g^{Le}_{mnkl}P^e_l\right]. \tag{18}$$

## III. RESULTS

The coefficients of the free-energy density function are found by fitting the temperature-dependent spontaneous polarization [10–13], and the optical properties and dynamic coefficients are fit to experimental data [14,15]. Specifically, we fit our model to the temperature-dependent refractive indices [14] and the dispersion of the refractive index [15], and then we predict the temperature-dependent electro-optic coefficients. The complete set of coefficients is given in table A1.

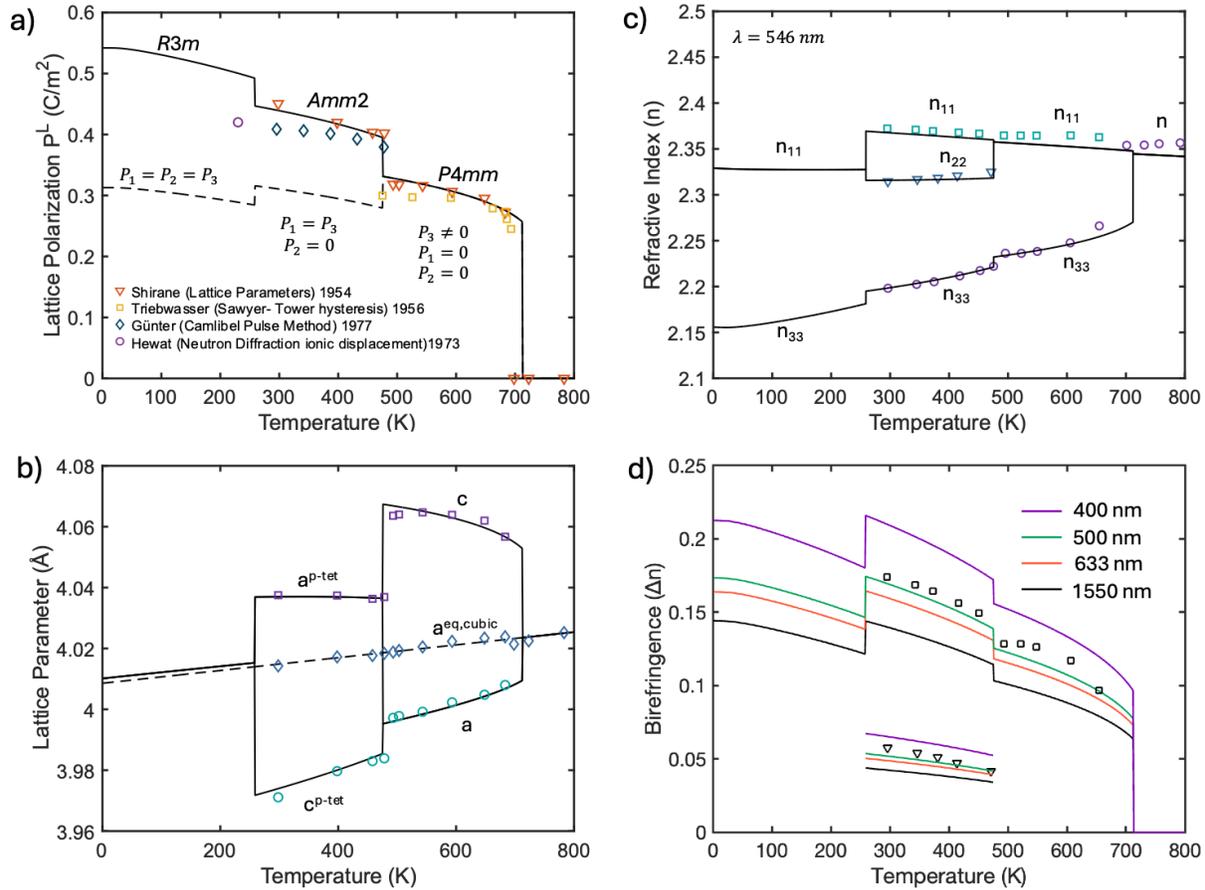

**Figure 1.** Temperature dependence of the spontaneous polarization, lattice parameters, refractive indices and birefringence. Solid lines denote the calculate result from thermodynamics, while symbols represent experimental values. (a) Calculated spontaneous lattice polarization as a function of temperature, compared with experimental values obtained from direct electrical measurements [11, 12] and values indirectly inferred from lattice parameter data [10,13]. (b) Temperature dependence of the lattice parameters showing close agreement with experimental measurements [10] (c) Temperature dependence of the refractive indices compared with experimental data from [14] (d) Temperature dependence of the birefringence compared with experimental data from [14]

Figure 1 illustrates the temperature dependence of the spontaneous polarization and refractive index. As shown in figure 1a, the calculated spontaneous polarization closely follows the experimental values obtained from directed electrical measurements and values indirectly inferred from lattice-parameter data (Appendix B). Upon cooling, $KNbO_3$ undergoes a series of ferroelectric phase transitions from cubic to tetragonal at 712 K, tetragonal to orthorhombic at 475 K, and orthorhombic to rhombohedral at 258 K. These ferroelectric phase transformations lead to discontinuities in the spontaneous polarization and reflect underlying changes to the crystal symmetry.

The same phase transitions are evident in the lattice parameters (figure 1b) and are also reflected as discontinuities in the temperature-dependent refractive index, as shown in figure 1c. In the tetragonal phase, KNbO$_3$ is a negative uniaxial crystal, indicating the presence of a single optical axis, which is parallel to the spontaneous polarization. The refractive indices perpendicular to this axis are equal, and the refractive index along the optic axis is smaller giving rise to the negative uniaxial behavior.

Upon entering the orthorhombic phase, it becomes a biaxial crystal, meaning that there are two unique optic axes before returning to a negative uniaxial crystal in the rhombohedral phase. The corresponding evolution of the birefringence, shown in figure 1d, further highlights these changes in symmetry. The correspondence between the spontaneous polarization and the refractive indices highlights the strong coupling between the ferroelectric order and the anisotropic optical properties.

Figure 2a shows the calculated lattice dielectric susceptibility compared to experimental measurements [16,17]. Overall, the thermodynamic model shows a strong agreement with the experimental results and falling within the variability expected for sample-to-sample differences. Near phase transitions the lattice dielectric permittivity is strongly enhanced which is reflected in the electro-optic behavior (Figure 3b). Near the cubic to tetragonal phase transition $\chi_{33}^L$ is enhanced, which leads to an enhanced $r_a^L$ electro-optic coefficient. As the temperature decreases, near the tetragonal to orthorhombic phase transition $\chi_{11}^L$ is enhanced which leads to a large $r_{131}^L$ and $r_{232}^L$ coefficient. In the orthorhombic phase $\chi_{22}^L$ increases, while $\chi_{11}^L$ decreases, leading to an increase in the $r_{232}^L$ coefficient and a decrease in the $r_{131}^L$ coefficient. The calculated temperature evolution and relative magnitudes of the electro-optic coefficients are in good agreement with experimental measurements [1,18] despite no direct fitting of the electro-optic effect. Minor quantitative disagreements, particularly in the $r_{232}^L$ electro-optic

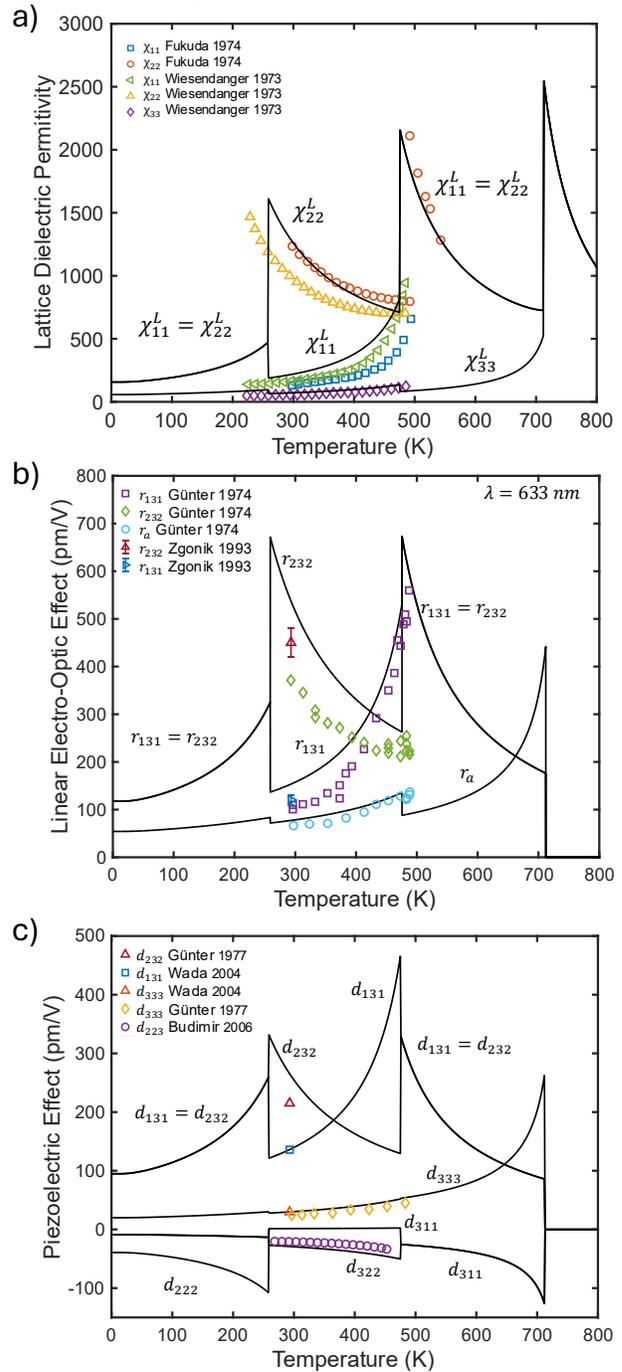

Figure 2. Temperature-dependent dielectric and electro-optic properties of KNbO$_3$ (a) Calculated lattice dielectric constant (solid line) compared with experimental data from [16,17], (b) Calculated temperature-dependent electro-optic effect at $\lambda = 633\ nm$ (solid line) compared with experimental data from [1, 18] (c) Temperature dependence of the piezoelectric coefficients compared with experimental measurements[19-21]

coefficients, are likely attributable to variations in measurement conditions between experiments. We further calculate the piezoelectric coefficients, which are also in good agreement with experimental measurements [19–21]. Notably, the temperature dependence of the piezoelectric coefficients (Figure 3c) mirrors the behavior of the electro-optic coefficients, with pronounced symmetry-dependent enhancements relating to the dielectric susceptibility.

**Table I.** Linear optical dielectric tensors, lattice dielectric tensors, electro-optic tensor, and piezoelectric tensors for the cubic, tetragonal, orthorhombic and rhombohedral phases.

| | Linear optical dielectric tensor $\lambda = 633nm$ | Linear lattice dielectric tensor | Electro-optic tensor (pm/V) $\lambda = 633nm$ | Piezoelectric tensor (pm/V) |
|---|---|---|---|---|
| Cubic Pm3m $P_i^L = (0,0,0)$ T=750K | $\begin{pmatrix} 5.32 & 0 & 0 \\ 0 & 5.32 & 0 \\ 0 & 0 & 5.32 \end{pmatrix}$ | $\begin{pmatrix} 1603 & 0 & 0 \\ 0 & 1603 & 0 \\ 0 & 0 & 1603 \end{pmatrix}$ | $\begin{pmatrix} 0 & 0 & 0 \\ 0 & 0 & 0 \\ 0 & 0 & 0 \\ 0 & 0 & 0 \\ 0 & 0 & 0 \\ 0 & 0 & 0 \end{pmatrix}$ | $\begin{pmatrix} 0 & 0 & 0 & 0 & 0 & 0 \\ 0 & 0 & 0 & 0 & 0 & 0 \\ 0 & 0 & 0 & 0 & 0 & 0 \end{pmatrix}$ |
| Tetragonal P4mm $P_i^L = (0,0,P)$ T=600K | $\begin{pmatrix} 5.36 & 0 & 0 \\ 0 & 5.36 & 0 \\ 0 & 0 & 4.90 \end{pmatrix}$ | $\begin{pmatrix} 971 & 0 & 0 \\ 0 & 971 & 0 \\ 0 & 0 & 140 \end{pmatrix}$ | $\begin{pmatrix} 0 & 0 & -5.50 \\ 0 & 0 & -5.50 \\ 0 & 0 & 133 \\ 0 & 280 & 0 \\ 280 & 0 & 0 \\ 0 & 0 & 0 \end{pmatrix}$ | $\begin{pmatrix} 0 & 0 & 0 & 0 & 136 & 0 \\ 0 & 0 & 0 & 136 & 0 & 0 \\ -40.3 & -40.3 & 83.7 & 0 & 0 & 0 \end{pmatrix}$ |
| Orthorhombic Amm2 $P_i^L = (P,0,P)$ T=293K | $\begin{pmatrix} 5.43 & 0 & 0 \\ 0 & 5.20 & 0 \\ 0 & 0 & 4.71 \end{pmatrix}$ | $\begin{pmatrix} 214 & 0 & 0 \\ 0 & 1307 & 0 \\ 0 & 0 & 70 \end{pmatrix}$ | $\begin{pmatrix} 0 & 0 & 18.3 \\ 0 & 0 & -4.00 \\ 0 & 0 & 73.1 \\ 0 & 540 & 0 \\ 153 & 0 & 0 \\ 0 & 0 & 0 \end{pmatrix}$ | $\begin{pmatrix} 0 & 0 & 0 & 0 & 136 & 0 \\ 0 & 0 & 0 & 265 & 0 & 0 \\ 1.37 & -29 & 29.9 & 0 & 0 & 0 \end{pmatrix}$ |
| Rhombohedral R3m $P_i^L = (P,P,P)$ T=200K | $\begin{pmatrix} 5.25 & 0 & 0 \\ 0 & 5.25 & 0 \\ 0 & 0 & 4.60 \end{pmatrix}$ | $\begin{pmatrix} 309 & 0 & 0 \\ 0 & 309 & 0 \\ 0 & 0 & 81 \end{pmatrix}$ | $\begin{pmatrix} 0 & 53.5 & 14.7 \\ 0 & -53.5 & 14.7 \\ 0 & 0 & 87.1 \\ 0 & 217 & 0 \\ 217 & 0 & 0 \\ 53.5 & 0 & 0 \end{pmatrix}$ | $\begin{pmatrix} 0 & 0 & 0 & 175 & 0 & 0 \\ 72.9 & -72.9 & 175 & 0 & 0 & 0 \\ -11.7 & -11.7 & 0 & 0 & 0 & 72.9 \end{pmatrix}$ |

**Table 1** summarizes the calculated property tensors for KNbO$_3$ in the cubic, tetragonal, orthorhombic and rhombohedral phases at different representative temperatures. The allowed tensor components reflect the crystallographic symmetry of each phase. In the cubic phase, the optical and lattice dielectric constant are isotropic, and the electro-optic and piezoelectric tensor vanish due to symmetry constraints. In the tetragonal phase, the polarization along the c-axis breaks the cubic symmetry leading to a uniaxial dielectric tensor and the appearance of nonzero electro-optic and piezoelectric coefficients. Notably, the large dielectric response perpendicular to the spontaneous polarization makes the $r_{131}$ and the $d_{131}$ electro-optic and piezoelectric coefficients larger compared to the other tensor components. In the orthorhombic phase, the polarization rotates to the [101]$_{pc}$ direction creating biaxial dielectric tensors with a distinct dielectric response along each principal direction. Here the $\chi_{22}^L$ response is enhanced while the $\chi_{11}^L$ response is diminished leading to a larger $r_{232}$ electro-optic coefficients compared to the larger $r_{131}$ coefficient. In the rhombohedral phase, the crystal is again uniaxial, and additional components of the electro-optic and piezoelectric coefficients are activated.

Figure 3 compares the calculated and experimentally measured anisotropic dispersion of the refractive index and electro-optic coefficients. Across the visible spectrum, the refractive indices exhibit a strong dispersion, which is well captured by equation 15. For simplicity we may drop the tensor notation, and express the electronic dielectric susceptibility as

$$\widetilde{\chi^e} = \tilde{n}^2 - 1 = \frac{1}{B^e - \epsilon_0(i\omega\gamma + \omega^2\mu)}. \quad (18)$$

In the thermodynamic limit ($\omega \to 0$), the frequency-dependent terms vanish and we find

$$\tilde{n}^2 - 1 = \frac{1}{B^e}. \quad (19)$$

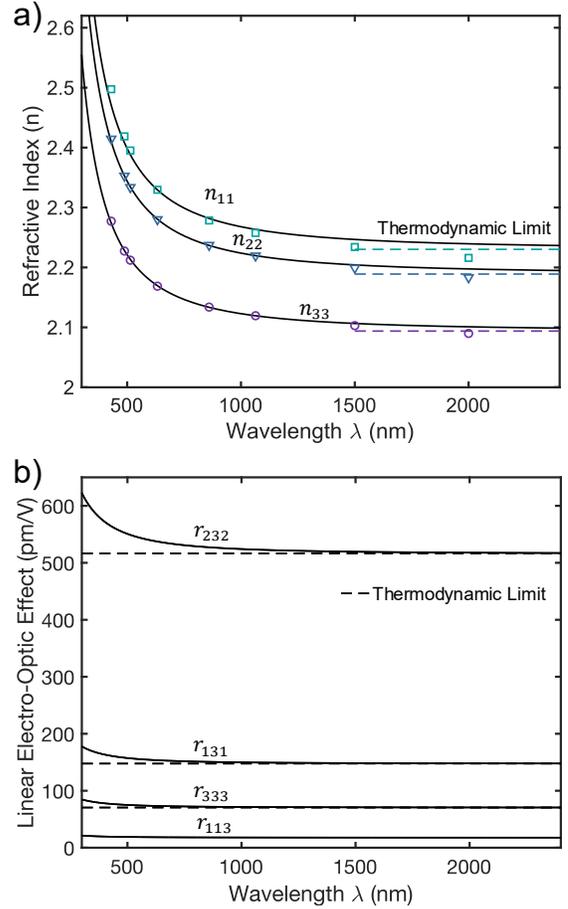

which corresponds to the dashed thermodynamic limit lines in figure 3a. Equation 18 shows that the optical dispersion is governed by the optical dielectric stiffness and the effective mass of the electronic polarization. Therefore, in the thermodynamic limit, when the refractive index is larger, there is a lower resonance frequency, and the onset of dispersion is pushed to longer wavelengths which is consistent with the experimental measurements.

Using equation 17 we further predict the dispersion of the electro-optic effect directly from the refractive index dispersion (Figure 3b). Again, for simplicity we may drop the tensor notation and ignoring absorption ($\gamma = 0$)

Figure 3. Wavelength-dependent optical properties in KNbO$_3$ (a) Dispersion of the refractive index (solid line) compared with experimental measurements [15] (markers) (b) dispersion of the electro-optic effect

$$r^L(\omega) = \frac{[2g^{LL}P^L + g^{Le}P^e]\chi^L}{(B^e + 1 - \epsilon_0\omega^2\mu^e)^2}. \quad (20)$$

As shown in figure 3b, the electro-optic coefficients remain close to their values in the thermodynamic limit over a broader wavelength range compared to the refractive index. This behavior follows from equation 20, where the effective resonance frequency is shifted to $B^e + 1$ to account for the vacuum dielectric susceptibility and the overall electro-optic dispersion is controlled by a squared denominator (e.g. $(B^e + 1 - \epsilon_0\omega^2\mu^e)^2$). This leads to the electro-optic coefficient staying in the thermodynamic limit for a broader frequency range, but it has a more dramatic dispersion as the resonance is approached.

## IV. DISCUSSION

The constructed thermodynamic model conceptually differs from previous Landau-Ginzburg-Devonshire models for KNbO$_3$ by including both the lattice and electronic polarization as coupled order parameters, thus connecting the ferroelectric and optical properties using a single free-energy expansion. Although the model is fitted based on a limited number of experiments, it is able to predict a wide range of anisotropic, temperature-dependent ferroelectric and optical properties across multiple phase transformations. The present study further highlights the utility of the thermodynamic theory of optical properties [9] for systematically describing the temperature- and symmetry-dependent optical properties in $KNbO_3$.

The thermodynamic model developed here provides the foundation for future studies aimed at engineering the optical properties of $KNbO_3$ using different elastic and electrostatic conditions. Furthermore, this framework will serve as the basis for modeling the optical properties in inhomogeneous crystals using the phase-field method [22]. In particular, the interactions between strain, domain structure and the electro-optic effect may be of interest to probe how nanoscale inhomogeneities modify the electro-optic response. Furthermore, lattice polarization dynamics and free carrier dynamics [26] can be readily incorporated, which opens the door to studying GHz-frequency electro-optic properties and photorefractive effects [27].

More broadly, $KNbO_3$ presents an alternative platform for electro-optic materials in contrast to $LiNbO_3$ and $BaTiO_3$. $KNbO_3$ has a 50% larger $g^{LL}_{1111}$ polar-optic coefficient than $BaTiO_3$ highlighting a stronger intrinsic coupling between the ferroelectric and optical properties. Furthermore it possesses multiple tunable phase transformations that can be controlled via strain [6], composition [5] and heterostructures. While the electro-optic response of $BaTiO_3$ excels at room temperature due to the proximity to the tetragonal to cubic phase transformation [9,28], $KNbO_3$ possesses the same sequence of phase transformations but they are all shifted to higher temperatures. Nevertheless, at room temperature, $KNbO_3$ is near the orthorhombic to rhombohedral phase transition temperature (figure 2b) making it a promising platform for future electro-optic modulators, which could be utilized for higher temperature applications and providing a broader window for tunable phase transformations.

## V. CONCLUSION

We have developed a new thermodynamic description of $KNbO_3$ based on the thermodynamic theory of optical properties, in which the lattice and electronic polarization are treated as coupled order parameters. Unlike existing Landau-Ginzburg-Devonshire descriptions that only describes the ferroelectric order, this formulation using a single free-energy function to capture the temperature dependent ferroelectric, dielectric, piezoelectric, and optical properties across multiple phase transformations. We fit the model using experimental measurements of the spontaneous polarization, dielectric susceptibilities, lattice parameters, and the refractive indices and predict the temperature-dependent electro-optic and piezoelectric effects which are all in close agreement with available experimental measurements. By incorporating the electronic polarization equation of motion, we further capture the dispersion of the refractive index and the electro-optic effect. More broadly, this work provides a consistent thermodynamic description of the ferroelectric and optical properties of $KNbO_3$, providing the foundation for future studies to investigate these properties in complex mesoscale domain structures using the phase-field method.

## ACKNOWLEDGEMNTS

The work was primarily supported by the National Science Foundation under award number DMR-2522897. A.R. acknowledges the support of the National Science Foundation Graduate Research Fellowship Program under Grant No. DGE1255832.

## DATA AVAILABILITY

The data that support the findings of this article are openly available [29].

## VI. REFERENCES

[1] P. Günter, Electro-optical properties of KNbO3, Optics Communications **11**, 285 (1974).
[2] T. Pliska, F. Mayer, D. Fluck, P. Günter, and D. Rytz, Nonlinear optical investigation of the optical homogeneity of KNbO$_3$ bulk crystals and ion-implanted waveguides, J. Opt. Soc. Am. B, JOSAB **12**, 1878 (1995).
[3] S. Triebwasser, Study of Ferroelectric Transitions of Solid-Solution Single Crystals of KNbO3-KTaO3, Phys. Rev. **114**, 63 (1959).
[4] H. Pohlmann, J.-J. Wang, B. Wang, and L.-Q. Chen, A thermodynamic potential and the temperature-composition phase diagram for single-crystalline K1-xNaxNbO3 (0 ≤ x ≤ 0.5), Applied Physics Letters **110**, 102906 (2017).
[5] B. Wang, H.-N. Chen, J.-J. Wang, and L.-Q. Chen, Ferroelectric domain structures and temperature-misfit strain phase diagrams of K1-xNaxNbO3 thin films: A phase-field study, Appl. Phys. Lett. **115**, 092902 (2019).
[6] S. Hazra et al., Colossal Strain Tuning of Ferroelectric Transitions in KNbO3 Thin Films, Advanced Materials **36**, 2408664 (2024).
[7] V. Gopalan and R. Raj, Domain structure and phase transitions in epitaxial KNbO3 thin films studied by in situ second harmonic generation measurements, Appl. Phys. Lett. **68**, 1323 (1996).
[8] L. Liang, Y. L. Li, L.-Q. Chen, S. Y. Hu, and G.-H. Lu, Thermodynamics and ferroelectric properties of KNbO3, J. Appl. Phys. **106**, 104118 (2009).
[9] A. Ross, M. S. M. M. Ali, A. Saha, R. Zu, V. Gopalan, I. Dabo, and L.-Q. Chen, Thermodynamic theory of linear optical and electro-optical properties of ferroelectrics, Phys. Rev. B **111**, 085109 (2025).
[10] G. Shirane, R. Newnham, and R. Pepinsky, Dielectric Properties and Phase Transitions of NaNb${\mathrm{O}}_{3}$ and (Na,K)Nb${\mathrm{O}}_{3}$, Phys. Rev. **96**, 581 (1954).
[11] S. Triebwasser, Behavior of Ferroelectric KNb${\mathrm{O}}_{3}$ in the Vicinity of the Cubic-Tetragonal Transition, Phys. Rev. **101**, 993 (1956).
[12] P. Günter, Spontaneous polarization and pyroelectric effect in KNbO3, J. Appl. Phys. **48**, 3475 (1977).
[13] A. W. Hewat, Cubic-tetragonal-orthorhombic-rhombohedral ferroelectric transitions in perovskite potassium niobate: neutron powder profile refinement of the structures, J. Phys. C: Solid State Phys. **6**, 2559 (1973).
[14] E. Wiesendanger, Optical properties of KNbO3, Ferroelectrics **1**, 141 (1970).
[15] B. Zysset, I. Biaggio, and P. Günter, Refractive indices of orthorhombic KNbO$_3$. I. Dispersion and temperature dependence, J. Opt. Soc. Am. B, JOSAB **9**, 380 (1992).
[16] T. Fukuda, H. Hirano, Y. Uematsu, and T. Ito, Dielectric Constant of Orthorhombic KNbO$_3$ Single Domain Crystal, Jpn. J. Appl. Phys. **13**, 1021 (1974).
[17] E. Wiesendanger, Dielectric, mechanical and optical properties of orthorhombic KNbO3, Ferroelectrics **6**, 263 (1973).
[18] M. Zgonik, R. Schlesser, I. Biaggio, E. Voit, J. Tscherry, and P. Günter, Materials constants of KNbO3 relevant for electro- and acousto-optics, J. Appl. Phys. **74**, 1287 (1993).
[19] P. Günter, Piezoelectric Tensor of KNbO3, Jpn. J. Appl. Phys. **16**, 1727 (1977).
[20] S. Wada, K. Muraoka, H. Kakemoto, T. Tsurumi, and H. Kumagai, Enhanced Piezoelectric Properties of Potassium Niobate Single Crystals by Domain Engineering, Jpn. J. Appl. Phys. **43**, 6692 (2004).
[21] M. Budimir, Piezoelectric Anisotropy and Free Energy Instability in Classic Perovskites, EPFL, 2006.
[22] J. H. Barrett, Dielectric Constant in Perovskite Type Crystals, Phys. Rev. **86**, 118 (1952).
[23] A. P. Levanyuk, B. A. Strukov, and A. Cano, Background dielectric permittivity: Material constant or fitting parameter?, Ferroelectrics **503**, 94 (2016).
[24] R. W. Boyd, *Nonlinear Optics* (Academic Press, 2020).
[25] L.-Q. Chen, Phase-Field Method of Phase Transitions/Domain Structures in Ferroelectric Thin Films: A Review, Journal of the American Ceramic Society **91**, 1835 (2008).

[26] T. Yang and L.-Q. Chen, Dynamical phase-field model of coupled electronic and structural processes, Npj Comput Mater **8**, 1 (2022).
[27] M. Zgonik, M. Ewart, C. Medrano, and P. Günter, *Photorefractive Effects in KNbO3*, in *Photorefractive Materials and Their Applications 2: Materials*, edited by P. Günter and J.-P. Huignard (Springer, New York, NY, 2007), pp. 205–240.
[28] A. R. Johnston, Dispersion of Electro-Optic Effect in BaTiO3, Journal of Applied Physics **42**, 3501 (1971).
[29] A. Ross, *Data and Codes for "Thermodynamics of Ferroelectric and Optical Properties in KNbO3,"* https://doi.org/10.5281/zenodo.18330243.

# Appendix A: Detailed expression of the free energy function and the associated parameters

For KNbO$_3$, we use the cubic phase as our high symmetry reference state and employ an 8th-order landau expansion to describe the relative stability of the lattice polarization compared to the cubic reference state. For this manuscript, we only consider the lowest-order polar-optic tensor, which is sufficient to describe the linear optical and electro-optic properties. Based on the symmetry of the reference phase we may expand equation 1 as

$$
\begin{aligned}
f(T, P_i^L P_i^e, E_i, \sigma_{ij}) &= f_o + a_{11}(T)[(P_1^L)^2 + (P_2^L)^2 + (P_3^L)^2] + a_{1111}[(P_1^L)^4 + (P_2^L)^4 + (P_3^L)^4] \\
&+ a_{1122}[(P_1^L)^2(P_2^L)^2 + (P_2^L)^2(P_3^L)^2 + (P_1^L)^2(P_3^L)^2] + a_{111111}[(P_1^L)^6 + (P_2^L)^6 + (P_3^L)^6] \\
&+ a_{111122}[(P_1^L)^4((P_2^L)^2 + (P_3^L)^2) + (P_2^L)^4((P_1^L)^2 + (P_3^L)^2) + (P_3^L)^4((P_1^L)^2 + (P_2^L)^2)] \\
&+ a_{112233}[(P_1^L)^2(P_2^L)^2(P_3^L)^2] + a_{11111111}[(P_1^L)^8 + (P_2^L)^8 + (P_3^L)^8] \\
&+ a_{11111122}[(P_1^L)^6((P_2^L)^2 + (P_3^L)^2) + (P_2^L)^6((P_1^L)^2 + (P_3^L)^2) + (P_3^L)^6((P_1^L)^2 + (P_2^L)^2)] \\
&+ a_{11112222}[(P_1^L)^4(P_2^L)^4 + (P_2^L)^4(P_3^L)^4 + (P_1^L)^4(P_3^L)^4] \\
&+ a_{11112233}[(P_1^L)^4(P_2^L)^2(P_3^L)^2 + (P_2^L)^4(P_1^L)^2(P_3^L)^2 + (P_3^L)^4(P_1^L)^2(P_2^L)^2] \\
&+ \frac{1}{2\epsilon_o} B_{11}^{ref}(T)[(P_1^e)^2 + (P_2^e)^2 + (P_3^e)^2] + \frac{1}{2\epsilon_o} g_{1111}^{LL}[(P_1^L)^2(P_1^e)^2 + (P_2^L)^2(P_2^e)^2 + (P_3^L)^2(P_3^e)^2] \\
&+ \frac{1}{2\epsilon_o} g_{1122}^{LL}[(P_1^L)^2((P_2^e)^2 + (P_3^e)^2) + (P_2^L)^2((P_1^e)^2 + (P_3^e)^2) + (P_3^L)^2((P_1^e)^2 + (P_2^e)^2)] \\
&+ \frac{2}{\epsilon_o} g_{1212}^{LL}[P_1^L P_2^L P_1^e P_2^e + P_2^L P_3^L P_2^e P_3^e + P_1^L P_3^L P_1^e P_3^e] \\
&- \frac{1}{2} s_{1111}(\sigma_{11}^2 + \sigma_{22}^2 + \sigma_{33}^2) - s_{1122}(\sigma_{11}\sigma_{22} + \sigma_{22}\sigma_{33} + \sigma_{11}\sigma_{33}) - 2 s_{1212}(\sigma_{12}^2 + \sigma_{23}^2 + \sigma_{13}^2) \\
&- Q_{1111}[\sigma_{11}(P_1^L)^2 + \sigma_{22}(P_2^L)^2 + \sigma_{33}(P_3^L)^2] \\
&- Q_{1122}[\sigma_{11}((P_2^L)^2 + (P_3^L)^2) + \sigma_{22}((P_1^L)^2 + (P_3^L)^2) + \sigma_{33}((P_1^L)^2 + (P_2^L)^2)] \\
&- Q_{1122}[\sigma_{11}((P_2^L)^2 + (P_3^L)^2) + \sigma_{22}((P_1^L)^2 + (P_3^L)^2) + \sigma_{33}((P_1^L)^2 + (P_2^L)^2)] \\
&- 2Q_{1212}[\sigma_{12} P_1^L P_2^L + \sigma_{23} P_2^L P_3^L + \sigma_{13} P_1^L P_3^L] \\
&- \frac{1}{2\epsilon_o} \pi_{1111}[\sigma_{11}(P_1^e)^2 + \sigma_{22}(P_2^e)^2 + \sigma_{33}(P_3^e)^2] \\
&- \frac{1}{2\epsilon_o} \pi_{1122}[\sigma_{11}((P_2^e)^2 + (P_3^e)^2) + \sigma_{22}((P_1^e)^2 + (P_3^e)^2) + \sigma_{33}((P_1^e)^2 + (P_2^e)^2)] \\
&- \frac{1}{\epsilon_o} \pi_{1212}[\sigma_{12} P_1^e P_2^e + \sigma_{23} P_2^e P_3^e + \sigma_{13} P_1^e P_3^e] \\
&- E_1 P_1^L - E_2 P_2^L - E_3 P_3^L \quad - E_1 P_1^e - E_2 P_2^e - E_3 P_3^e \; -\epsilon_0(E_1^2 + E_2^2 + E_3^2).
\end{aligned}
\tag{A1}
$$

where the coefficients used for this paper are given in Table A1.

**Table A1.** Coefficients in the thermodynamic free energy function and equation of motion

| | | | |
|---|---|---|---|
| $g_{1111}^{LL}$ | $27.8 \times 10^{-2} (m^4/C^2)$ | $p_{1111}$ | 0.5328 (Unitless) |
| $g_{1122}^{LL}$ | $-1.1 \times 10^{-2} (m^4/C^2)$ | $p_{1122}$ | 0.1584 (Unitless) |
| $g_{1212}^{LL}$ | $8.2 \times 10^{-2} (m^4/C^2)$ | $p_{1212}$ | $-0.432$ (Unitless) |
| $a_{11}$ | $a_o \left( T_s \coth\left(\frac{T_s}{T}\right) - T_s \coth\left(\frac{T_s}{T_c}\right) \right)$ | $\mu_e$ | $32.8 \times 10^{-23} \left(\frac{Kg}{m} \frac{m^4}{C^2}\right)$ |
| $T_c$ | $650K$ | $\gamma_e$ | $3 \times 10^{-9} \left(\frac{Kg}{ms}\frac{m^4}{C^2}\right)$ |
| $T_s$ | $54K$ | $B_{ij}^{e,ref}(T_0)$ | $0.2546$ (Unitless) |
| $a_0$ | $3.53 \times 10^5 \left(\frac{J}{m^3}\frac{m^4}{KC^2}\right)$ | $T_0$ | $398K$ |
| $a_{1111}$ | $-5.86 \times 10^8 \left(\frac{J}{m^3}\frac{m^8}{C^4}\right)$ | $\alpha_{11}$ | $2.657 \times 10^{-5} \left(\frac{1}{K}\right)$ |
| $a_{1122}$ | $9.66 \times 10^8 \left(\frac{J}{m^3}\frac{m^8}{C^4}\right)$ | $a_{11111111}$ | $1.74 \times 10^{10} \left(\frac{J}{m^3}\frac{m^{16}}{C^8}\right)$ |
| $a_{111111}$ | $2.71 \times 10^9 \left(\frac{J}{m^3}\frac{m^{12}}{C^6}\right)$ | $a_{11111122}$ | $5.99 \times 10^9 \left(\frac{J}{m^3}\frac{m^{16}}{C^8}\right)$ |
| $a_{111122}$ | $-2.2 \times 10^9 \left(\frac{J}{m^3}\frac{m^{12}}{C^6}\right)$ | $a_{11112222}$ | $2.5 \times 10^{10} \left(\frac{J}{m^3}\frac{m^{16}}{C^8}\right)$ |
| $a_{112233}$ | $4.4 \times 10^9 \left(\frac{J}{m^3}\frac{m^{12}}{C^6}\right)$ | $a_{11112233}$ | $-1.63 \times 10^{10} \left(\frac{J}{m^3}\frac{m^{16}}{C^8}\right)$ |
| $Q_{1111}$ | $0.11 \left(\frac{m^4}{C^2}\right)$ | $Q_{1122}$ | $-0.053 \left(\frac{m^4}{C^2}\right)$ |
| $Q_{1212}$ | $0.026 \left(\frac{m^4}{C^2}\right)$ | | |

**APPENDIX B: Calculation of polarization in KNbO$_3$ From Lattice Parameters**

The lattice polarization is related to the spontaneous strain through

$$\varepsilon_{ij}^o = Q_{ijkl} P_k^L P_l^L, \tag{B1}$$

where $\varepsilon_{ij}^o$ is the spontaneous strain with respect to the equivalent paraelectric reference state.

In the tetragonal phase, we may write where under zero-electric field $P_1^L = P_2^L = 0$ and $P_3^L \neq 0$ we may write

$$\varepsilon_{33}^o = \frac{c - a^o}{a^o} = Q_{1111}(P_3^L)^2, \tag{B2}$$

$$\varepsilon_{11}^o = \varepsilon_{22}^o = Q_{1122}(P_3^L)^2 = \frac{a - a^o}{a^o}, \tag{B3}$$

where $c$ and $a$ are the lattice parameters of the tetragonal ferroelectric phase and $a^o$ is the equivalent lattice parameters of the paraelectric phase. We may combine equations B2 and B3 to find

$$z = \frac{c}{a} = \frac{Q_{1111}(P_3^L)^2 + 1}{Q_{1122}(P_3^L)^2 + 1}, \tag{B4}$$

where $z$ is the relative c/a ratio. We may rearrange equation B4 to find the spontaneous polarization

$$P_3^L = \left( \frac{z - 1}{Q_{3333} - zQ_{1133}} \right)^{1/2}. \tag{B5}$$

For the orthorhombic phase, under zero electric field $P_1^L = P_2^L \neq 0$ and $P_3^L = 0$ we may use the pseudo-tetragonal unit cell convention and write

$$\varepsilon_{33}^o = \frac{c - a^o}{a^o} = 2Q_{1122}(P_1^L)^2, \tag{B6}$$

$$\varepsilon_{11}^o = \varepsilon_{22}^o = \frac{a - a^o}{a^o} = (Q_{1111} + Q_{1122})(P_1^L)^2. \tag{B7}$$

We may combine equations B6 and B7 to find

$$z = \frac{c}{a} = \frac{2Q_{1122}(P_1^L)^2 + 1}{(Q_{1111} + Q_{1122})(P_1^L)^2 + 1} \tag{B8}$$

where $z$ is the relative c/a ratio. We may rearrange equation B8 to find the spontaneous polarization

$$P_1^L = \left( \frac{z - 1}{2Q_{1122} - z(Q_{1111} + Q_{1122})} \right)^{1/2}. \tag{B9}$$